\documentclass[final,5p,times,twocolumn]{elsarticle}
\usepackage{amsmath,amssymb,amsfonts,scalerel}
\usepackage{graphicx}
\usepackage{subcaption}
\usepackage{epsfig}
\usepackage{float}
\usepackage{xcolor}
\usepackage[colorlinks=true,citecolor=blue,linkcolor=blue,urlcolor=blue]{hyperref}
\usepackage{setspace}
\usepackage{multirow}
\usepackage{multicol}
\usepackage{url}
\usepackage{rotating}
\usepackage{geometry}
\biboptions{sort,compress}
\journal{Physica~B:~Physics~of~Condensed~Matter}

\begin{document}
\begin{frontmatter}

\title{Dopant~size~effect~on~BiFeO$\rm_{3}$~perovskite\
       structure~for~enhanced~photovoltaic~activity}
\author{T.E.~Ada}
\ead{tewodros.eyob@aau.edu.et} 
\author{K.N.~Nigussa\corref{cor1}}
\cortext[cor1]{Corresponding author:\ kenate.nemera@aau.edu.et\ 
               (K.N.~Nigussa)}
\author{L.D.~Deja}
\ead{lemi.demeyu@aau.edu.et}

\address{Department of Physics,\ Addis Ababa University,\ 
         P.O. Box 1176,\ Addis Ababa,\ Ethiopia}
\begin{abstract}
This\ study\ is\ carried\ out\ using\ first\
principles\ density\ functional\ theory\ 
calculations\ within\ gpaw\ code.\ Atomic\ 
size\ effect is\ analyzed and\ investigated\ 
by\ doping either Li,\ Cs\ or both\ on Barium\ 
doped\ BiFeO$\rm_{3}$~(BFO)\ which\ belongs\ 
to\ monoclinic\ $P2\rm_{1}/m$\ space\ group.\ 
The calculated\ results reveal\ that\ Cs\ 
doped BFO\ had significantly\ improved\ 
photocurrent\ density\ seemingly\ due to\ 
broadened\ absorption peaks $\&$\ biplasmons\ 
generation.\ Co-doping\ two atoms\ 
with large size\ difference have\ a\ 
significant\ effect\ on plasmon\ width\ 
$\&$\ peak\ than doping\ with\ a\ single\ 
$\&$\ small sized atom.\ 
At\ higher\ photon\ energy realms\ of\ 
the\ order\ 10~eV,\ the\ index of\ 
refraction\ reduces\ to\ 
$n(\omega)~{\rightarrow}~1$~implying\ 
that light\ wave\ can\ tunnel\ through\ 
the\ pristine\ $\&$\ doped BFO\ without\ 
any\ phase change,\ thus\ indicating\ its\ 
potential\ as\ an\ efficient\ candidate\ 
for\ a\ photonic\ application.\ In addition,\ 
doped BFO\ shows abundant\ photocurrent\ 
generation\ properties\ which\ would be\ 
important\ in\ solar\ cell\ $\&$\ photovoltaic\ 
applications.\ 
\end{abstract}


\end{frontmatter}

\section{Introduction\label{sec:intro}}
BiFeO$\rm_{3}$~(BFO)\ is multiferroic perovskite\ 
crystal structure,\ which permits\ coexistence of\  
varies\ orders at\ room temperature.\ Doping fine\ 
or\ large sized\ atoms further\ improves\ simultaneous\ 
coexistence\ of different\ orders in\ single phase\ 
resulting in\ enhanced structural,\ optical\ and\ 
electronic\ properties\ of materials~\cite{YKTNS2001}.\

Several\ attempts\ have been done\ in\ 
advancement\ of the\ ferromagnetic\ properties\ 
of BFO\ $\&$\ for\ improvements\ in leakage\ current\ 
amount\ decrement\ by\ doping\ either\ in A-site\ 
or\ B-site\ or\ both sites~\cite{YXSKAOW2014}.\ 
Shaan Ameer~et~al~\cite{AJTJG2018}\ reported that\ 
doping Li\ on BFO\ reduces ferromagentism.\ However,\ 
optical property\ enhanced due to small optical\ 
band gap energy. Moreover,\ Aungkan Sen~et~al~\cite{SHIHZMG2020}\
suggested\ proportional\ doping\ of\ Ba\ $\&$\ Mo\ 
in\ BFO\ result in\ improved optical\ absorbance\  
and\ dielectric\ property.

Highest level\ of doping\ not only\ alter the\ 
properties of\ the materials,\ but had remarkable\ 
impact on\ the crystal\ structure as well.\ For\ 
instance,\ La-doped\ BFO structure\ is\ transformed\ 
from\ rhombohedral\ to\ triclinic\ or\ orthorhombic\ 
around\ 10\ mol $\%$,\ and a\ smaller\ ionic\ radius\ 
rare-earth\ doping at\ the\ B-site\ causes\ structural\ 
distortion,\ and\ brought\ forth\ antiferroelectric-like\ 
double-hysteresis\ P-E loops.\ Moreover,\ rare-earth\ 
elements\ with a\ smaller\ ion\ radius,\ because\ of\ 
a\ geometrical\ $\&$\ magnetic moments\ factor,\ are\ 
forced to\ align themselves\ either in\ symmetrical\ 
or antisymmetrical\ orientation,\ resulting in\ various\ 
magnetic\ order phases.\ Arnold~et~al~\cite{Arnold2015}\ 
also\ ascertained\ that\ rare-earth\ doped BFO materials'\ 
ionic radii\ at B-site\ are the\ prominent source\ 
for phase\ transitions and\ structural phase\ 
transitions, which\ is\ also\ supported\ by\ 
other\ literatures~\cite{TAO201783, NG2008, KPACFNRT2010}.\ 

Besides\ phase,\ structural\ and property\ alteration,\  
chemical\ strain\ forces can\ be\ exhibited\ among\ 
atoms\ because of\ short and\ long range\ interactions.\ 
J.A.~Schiemer et al~\cite{SWLC2013}\ studied\ 
significance of\ strain in\ coupling phase orders\  
with effect\ on elastic\ response of either in\ 
high or low\ level doped\ crystals in\ relation\ 
to\ magnetic order,\ oxygen vacancy\ dynamics,\ 
and\ conductivity.\ It\ is\ also inferred that\ 
important\ transformation\ in multiferroic\ 
properties are\ highly correlated\ to structural\ 
changes.\ It has been\ well understood\ that\ 
applied strain\ on BFO scales\ up\ the magnetic,\ 
piezoelectric,\ ferroelectric,\ and optical\ 
properties~\cite{HOIA2014}.\

Liang\ Bai~et~al~\cite{BSMYZL2020}\ also\ 
experimented\ and observed\ that\ doping\ 
Co$\rm^{3+}$\ ions\ knocks out\ Fe$\rm^{3+}$\ 
ions from\ B-sites\ and\ opens\ up\ for\ 
random\ substitute.\ Consequently,\ a\ spirally\ 
arranged\ spin becomes\ disrupted\ and\ 
chemical\ strain\ is\ grown\ because of\ 
uneven\ size of\ the two\ B-site\ cations\ 
(Co$\rm^{3+}$ and Fe$\rm^{3+}$\ ions)).\
In spite of\ the previous efforts,\ strain\ 
evolution due\ to dopant size\ $\&$\ its\ 
facilitation for\ coupling of\ multiferroic\ 
orders still\ remains\ debatable.\ 
Here we\ considered\ perovskite monoclinic\ 
BFO~($P2\rm_{1}/m$)\ phase $\&$ the\ experimentally\ 
synthesized\ counterpart\ as\ heated over\ 
a\ range of high\ temperature\ during sintering\ 
processes,\ is\ evidenced with\ appearance\ 
of several\ phases plus\ parasitic\ phase\
(Bi$\rm_{2}$Fe$\rm_{4}$O$\rm_{9}$)~\cite{HKLBKD2008}.\ 

\begin{figure*}[ht!]
    \begin{subfigure}[t]{0.5\textwidth}
        \centering
        \includegraphics[height=2.75in]{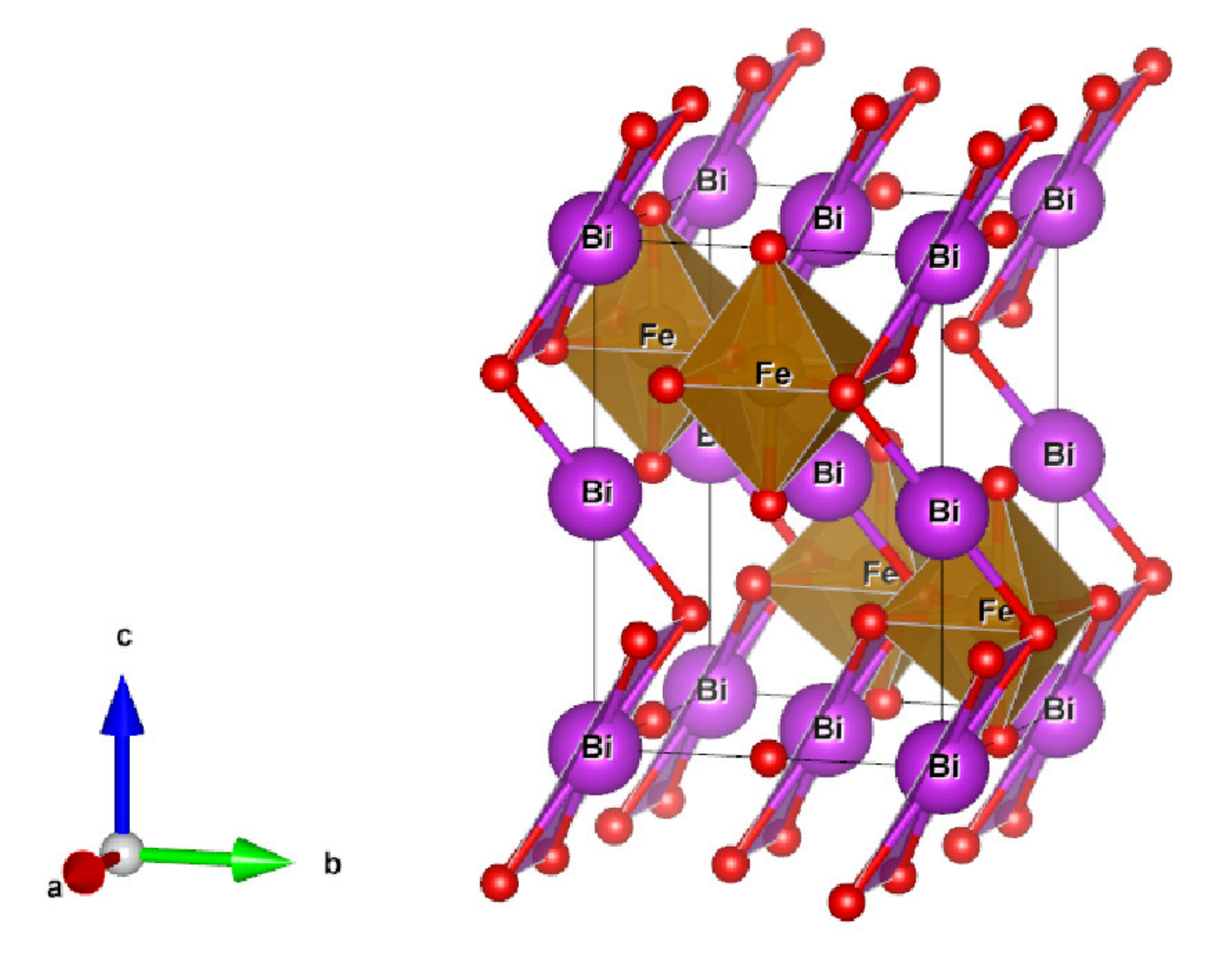}
        \caption{\label{fig1a} Pristine BiFeO$\rm_{3}$~(P2$\rm_{1}/m$)}
    \end{subfigure}%
    ~ 
     \begin{subfigure}[t]{0.5\textwidth}
        \centering
        \includegraphics[height=2.75in]{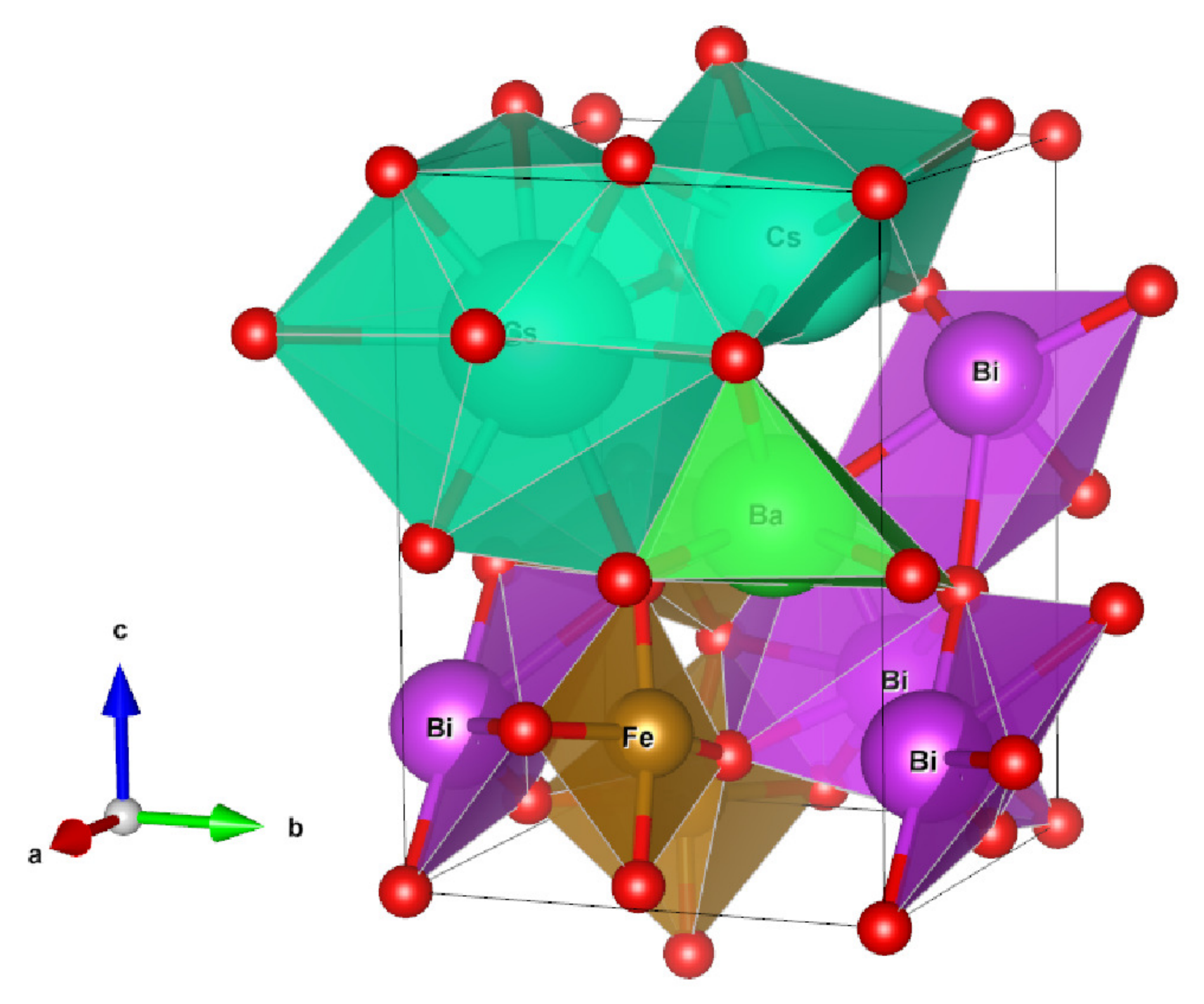}
        \caption{\label{fig1b} Li, Cs co-doped BiFeO$\rm_{3}$~(P2$\rm_{1}/m$)}
    \end{subfigure}
    ~
    \begin{subfigure}[t]{0.5\textwidth}
        \centering
        \includegraphics[height=2.75in]{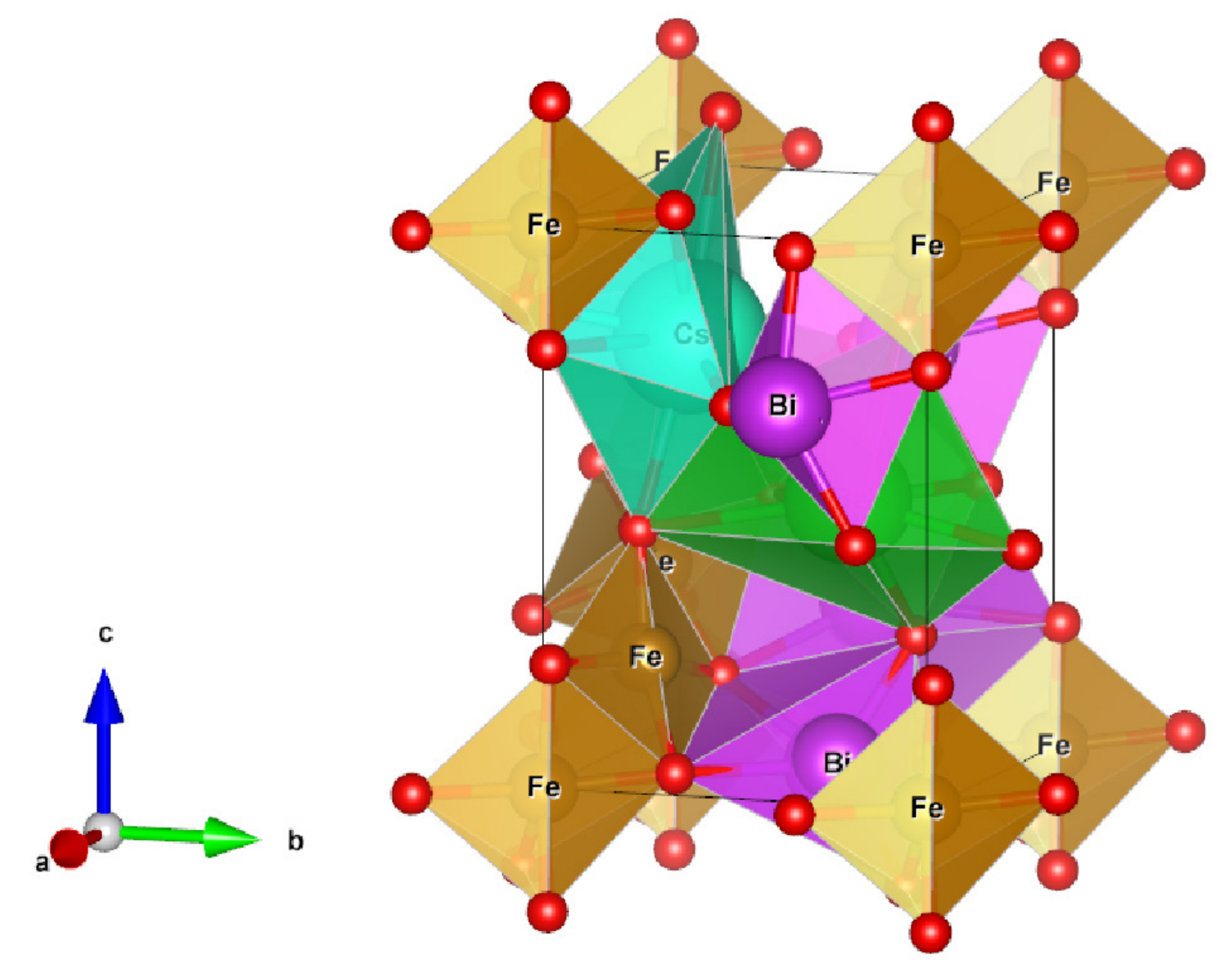}
        \caption{\label{fig1c} Li doped BiFeO$\rm_{3}$~(P2$\rm_{1}/m$)}
    \end{subfigure}
    ~
     \begin{subfigure}[t]{0.5\textwidth}
        \centering
        \includegraphics[height=2.75in]{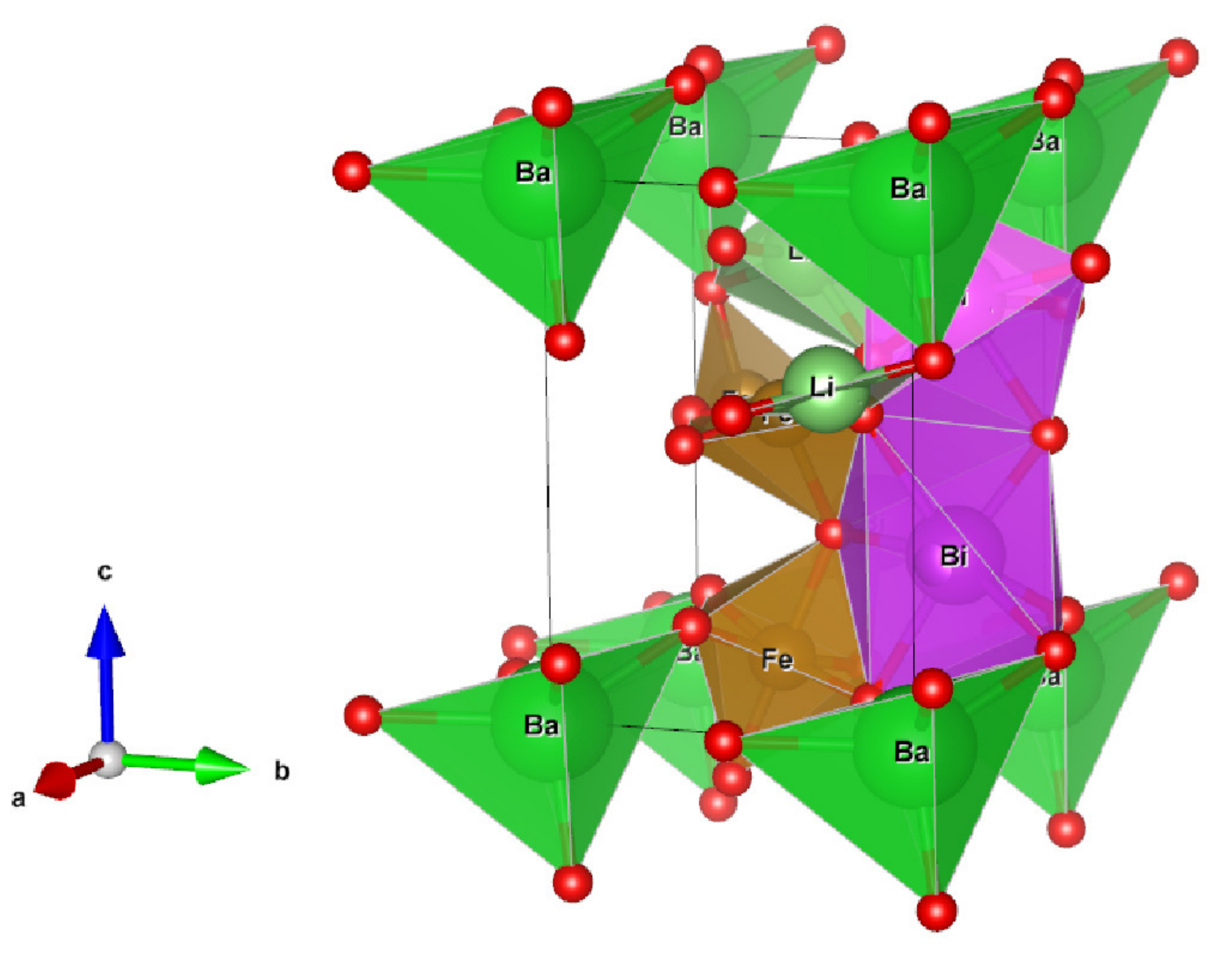}
        \caption{\label{fig1d} Cs doped BiFeO$\rm_{3}$~(P2$\rm_{1}/m$)}
    \end{subfigure}%
    \caption{\label{fig1} Color\ online.~Crystal structure of\ 
             bulk pristine\ BiFeO$\rm_{3}$ in\ comparison with\
             doped\ BiFeO$\rm_{3}$\ illustrating\ the structural\ 
             tilting of\ BiO$\rm_{6}$ polyhedra\ relative to\ 
             neighbouring~polyhedrons.}   
\end{figure*}

In the\ course of sintering\ process,\ various\ 
transformation\ is\ undergone\ as function of\ 
temperature\ leading to\ bismuth depreciated,\ 
due to its\ volatility resulting\ in impure BFO\ 
phases,\ which is\ often called\ secondary phase\ 
of\ the\ cubic structure.\ However,\ these phases\ 
are\ further purified\ by proportional\ addition\ 
of\ Bi$\rm_{2}$O$\rm_{3}$\ and\ 
Fe$\rm_{2}$O$\rm_{3}$~\cite{NS2017},\ 
see Eq.~(\ref{eq1}).
\begin{equation}
Bi\rm_{2}Fe\rm_{4}O\rm_{9}+Bi\rm_{2}O\rm_{3} \rightarrow 4BiFeO\rm_{3}
\label{eq1}
\end{equation}  
We\ considered\ perovskite monoclinic\ BFO~($P2\rm_{1}/m$)\ 
phase because\ one could\ easily understand\ doping\ 
effect.\ Therefore,\ this work\ explores\ cesium~(Cs)\ 
and lithium~(Li)\ size\ effect\ on barium~(Ba)\ 
doped\ BiFeO$\rm_{3}$.\ It results\ in structural\ 
phase transition\ due to strain\ created because\ 
of size of\ substitute atoms~(Fig.~\ref{fig1}).\
%%%%%%%%%%%Figure 1%%%%%%%%%%%

%%%%%%%%%%%%%%%%%%%%%%%%%%  
\section{Computational\ method\label{sec:compmeth}}
Ab initio\ calculations\ were\ performed using\ the\ 
projector\ augmented\ wave~(PAW)\ method as implemented\ 
in gpaw code~\cite{Enkovaaraetal2010}.\ The electron\ 
wave-function\ is approximated\ using the\ implementation\ 
of a\ projector augmented\ wave method~\cite{MHJ2005}\ 
$\&$\ the Schr$\rm\ddot{o}$dinger\ equation applied to\ 
total energy\ functions are\ solved self-consistently\ 
using the\ Kohn-Sham\ scheme~\cite{KS65}.\ All electron\ 
waves are\ expanded over\ periodic potential\ plane\ 
with a method\ of P.E.~Bl$\rm\ddot{o}$chl~\cite{Blochl94}\ 
having\ a band,\ a k-points,\ and a\ reciprocal\ 
lattice\ grid indices.\ A plane\ wave cut-off\ energy\ 
used is\ 520 eV.\ The k-points\ within\ Brillouin\ 
zone~(BZ)~is\ chosen\ according to\ Monkhorst-Pack\ 
scheme~\cite{MP76},\ where a\ $\bf{k}$-mesh\ of\ 
8$\times$8$\times$8\ is used.\

The\ interactions\ of the valence\ electrons\ 
with\ the\ core electrons\ and nuclei\ is\ 
treated\ within\ a projector\ augmented wave\ 
(paw) data\ sets~\cite{Enkovaaraetal2010,MHJ2005}.\
The number of\ valence electrons\ considered\ 
for\ each\ element within\ the paw\ data sets\ 
is\ Bi~(5d$\rm^{10}$6s$\rm^{2}$6p$\rm^{3}$),\ 
Fe~(3d$\rm^{6}$4s$\rm^{2}$),\ 
O~(2s$\rm^{2}$2p$\rm^{4}$),\
Cs~(5s$\rm^{2}$5p$\rm^{6}$6s$\rm^{1}$),\
Li~(2s$\rm^{1}$),\ $\&$\
Ba~(5s$\rm^{2}$5p$\rm^{6}$6s$\rm^{2}$),\
Geometry\ relaxations\ are carried out\ using BFGS\ 
minimizer~\cite{BS82},\ where\ optimizations\ 
of the\ atomic coordinates\ and\ the unit cell\ 
degrees of\ freedom is\ done within\ the concept\ 
of the\ Hellmann-Feynman\ forces and\ 
stresses~\cite{RF39,NM85}\ calculated on\ 
the Born-Oppenheimer~(BO)~surface~\cite{WM91}.\ 
The convergence\ criteria for\ the forces\ 
were set\ at 0.05 $\rm{eV/\AA}$.\ The\ 
exchange-correlation\ energies are\ approximated\ 
within the\ generalized\ gradient approximation\ 
of PBE~\cite{PBE96}\ $\&$\ a $\bf{k}$-mesh of\ 
8$\times$8$\times$8\ is used\ in a geometry\ 
relaxation calculations.\ The\ strongly\ 
correlated\ nature\ of\ $d$\ electrons\ 
of\ Fe\ were\ treated\ using\ Hubbard-like\ 
model~which is\ introduced into the Gpaw\ 
code according to~\cite{LAZ95,DBSHS98},\ 
where\ U-J=7.5~eV.\ 

Brillouin zone\ integration were performed\ 
with the tetrahedron\ method which is proved\ 
efficient especially\ for excited states\ $\&$\ 
magnetically induced\ dielectric function\ 
calculation~\cite{MVC79}.\ Ground state\ 
calculation is\ performed with k-point\ 
density of 5 points per $\rm\AA^{-1}$ with\ 
$\Gamma$-point inclusion.~A restarted\ 
calculations\ were\ done with\ a fixed\ 
density of 7.5\ points per $\rm\AA^{-1}$\ 
over densely\ sampled k-points grid.\ 
The k-point\ density is\ calculated as:\
\begin{equation}
N \frac{\bf{a\rm^{*}}}{2\pi}
\end{equation}
where N\ is the number\ of k-points and\ 
$\bf{a\rm^{*}}$\ is the\ reciprocal\ of\ 
a\ lattice vector\ $\bf{a}$\ of\ the\ 
unit-cell.\ 
One can\ analyse\ and\ understand~material's\ 
optical\ response\ property\ once\ incident\ 
photon\ energy is\ imparted\ to~the electron,\ 
where\ the\ material's\ response or\ dielectric\ 
function\ is given as\
\begin{equation}
\varepsilon(\omega)=\varepsilon\rm_{1}
(\omega)+i\varepsilon\rm_{2}(\omega)
\label{eq4}
\end{equation}
The imaginary part\ $\varepsilon\rm_{2}(\omega)$\ 
is\ calculated\ from the density\ matrix of the\ 
electronic\ structure~\cite{HL87}\ according\ to\ 
the\ implementations\ by\ the\ group\ of\ 
G.\ Kresse~\cite{GHKFB2006, SK2006},\ 
$\&$\ given by\
\begin{equation}{\label{eq5}}
\varepsilon\rm_{2}(\omega)=\frac{2e^{2}}
{\Omega {{\omega}^{2}}{m_{e}^{2}}} 
{\sum\limits_{k,v,c}}{w\rm_{k}}{{\mid}
\langle{\psi\rm_{k}^{c}}{\mid}{\bf u}{\cdot}
{\bf r}{\mid}{\psi\rm_{k}^{v}}\rangle{\mid}}^{2}
\delta(E\rm_{k}^{c}-E\rm_{k}^{v}-\hbar \omega), 
\end{equation}
where $e$ is the\ electronic charge,\ and\ 
$\psi\rm_{k}^{c}$\ and\ $\psi\rm_{k}^{v}$\ are\ 
the conduction band\ (CB)\ and\ valence band\ 
(VB)\ wave functions at k,\ respectively,\ 
$\hbar \omega$\ is the\ energy of the\ incident\ 
phonon,\ ${\bf u}{\cdot}{\bf r}$\ is\ the\ 
momentum operator,\ $w\rm_{k}$\ is a joint\ 
density of states,\ $\&$\ $\Omega$\ is\ volume\ 
of\ the\ primitive cell.\

The\ Real\ part\ $\varepsilon\rm_{1}(\omega)$\ 
of\ the dielectric\ function\ can be\ found\ 
from\ the\ Kramer-Kronig\ equation~\cite{FW72}.\
\begin{equation}
\varepsilon\rm_{1}(\omega)=1+\frac{2}{\pi}
P\int_{0}^{\infty}\frac{\omega^{'}
\varepsilon\rm_{2}(\omega^{'})d\omega^{'}}
{\omega^{'2}-\omega^{2}}
\end{equation}
where,\ P stands for\ the principal\ value\ 
of\ the integral.\ 
The optical\ absorption\ coefficient\ 
was\ obtained by\ using Eq.~(\ref{eq6})
\begin{equation}
\alpha=\sqrt{2}\frac{\omega}{c}
\sqrt{\sqrt{\varepsilon\rm^{2}_{1}
(\omega)+\varepsilon\rm^{2}_{2}
(\omega)}-\varepsilon\rm_{1}(\omega)}
\label{eq6}
\end{equation}
where\ $\omega$ is\ photon\ frequency,\ 
$c$\ is speed\ of light.\ 
$\varepsilon\rm_{1}$ $\&$ $\varepsilon\rm_{2}$\ 
are\ frequency dependent\ real and imaginary\ 
parts of\ dielectric function\ as stated in\ 
Eq.~(\ref{eq4}).\ For photovoltaic\ applications,\ 
the BFO should\ be\ able to absorb\ as much\ 
light as\ possible\ to\ generate a\ photocurrent,\ 
which requires\ a\ lower band\ gap\ and a\ large\ 
absorption\ coefficient.\

The photocurrent\ density under irradiation\ 
with a certain\ wavelength\ can be described\ 
by an\ empirical\ Glass Law~\cite{GLN74}.
\begin{equation}
J\rm_{ph}=\alpha \kappa I
\label{eq7}
\end{equation}
where\ $\alpha$ is\ the optical absorption\ 
coefficient,\ $\kappa$\ is a\ material-dependent\ 
Glass\ coefficient,\ and I is\ irradiation\ intensity.\ 
The photocurrent\ density\ $J\rm_{ph}$ is\ linearly\ 
proportional\ to the absorption\ coefficient.\ The\ 
relation between\ $\alpha$\ and an optical\ band gap\ 
$E\rm_{g}^{o}$\ can be estimated\ using the Tauc\ 
relation~\cite{TGV66}.\
\begin{equation}
\alpha({\hbar}{\omega})\sim A({\hbar}
{\omega}-E\rm_{g}^{o})^{\frac{1}{2}}
\label{eq8}
\end{equation}
where A\ is a material\ dependent constant and\ 
${\hbar}{\omega}$\ is the\ incident photon energy.\ 
According to\ Eq.~(\ref{eq7})\ $\&$\ Eq.~(\ref{eq8}),\ 
one can\ extrapolate a\ larger photocurrent\ density\ 
by\ increasing the\ absorption coefficient\ with a\ 
small band gap.\ 

From\ dielectric\ function,\ all the other\ 
optical\ properties\ such as,\ reflectivity\ 
$R$,\ refractive\ index\ $n\rm_{0}$,\ $\&$\ 
extinction\ coefficient\ $\kappa$\ is\ 
also\ obtained~\cite{TKL2021, SHDYLW2015, TJCMZ2018},\ 
$\&$\ given as follows.\ 
\begin{equation}
n(\omega)=\frac{1}{\sqrt{2}}
\sqrt{\sqrt{\varepsilon\rm_{1}^{2}
(\omega)+\varepsilon\rm_{2}^{2}
(\omega)}+\varepsilon\rm_{1}(\omega)}
\label{eq9}
\end{equation}
In addition,\ reflectivity\ $R(\omega)$,\ $\&$\ energy\ 
loss\ function\ $\iota(\omega)$\ is\ calculated\ 
as\ 
\begin{equation}
R(\omega)=\left|\frac{\sqrt{\varepsilon(\omega)}-1}
{\sqrt{\varepsilon(\omega)}+1}\right|^{2}
\end{equation}
and
\begin{equation}
\iota(\omega)=\frac{\varepsilon\rm_{2}(\omega)}
{{\Bigg[}\varepsilon\rm_{1}^{2}(\omega)+
\varepsilon\rm_{2}^{2}(\omega){\Bigg]}},
\end{equation}
respectively.\ 

The spin-orbit\ module calculates\ spin-orbit\ 
band structures\ non-self consistently.\ The input\ 
is a standard\ converged GPAW\ calculation and the\
module diagonalizes\ the spin-orbit Hamiltonian\ 
in a basis of\ scalar-relativistic\ Kohn-Sham\ 
eigenstates.\ Since the spin-obit\ coupling is\ 
largest\ close to the nuclei,\ we only consider\ 
contributions from\ inside the PAW\ augmentation\ 
spheres where\ the all-electron\ states can be\ 
expanded as\

\begin{equation}
|\psi\rm_{nk}\rangle=\sum\rm_{ai}\langle\tilde 
p\rm_i^a|\tilde\psi\rm_{nk}\rangle|\phi\rm_i^a\rangle    
\end{equation}

The full Bloch\ Hamiltonian in\ a basis of scalar\ 
relativistic\ states becomes\

\begin{equation}
\begin{array}{cc}
H\rm_{nn'\sigma\sigma'}(k)=\varepsilon\rm_{nk\sigma}
\delta\rm_{nn'\sigma\sigma'}+\langle\psi\rm_{nk\sigma}
|H^{SO}(k)|\psi\rm_{n'k\sigma'}\rangle= 
\varepsilon\rm_{nk\sigma}\delta\rm_{nn'\sigma\sigma'}\\
+\sum\limits_{i\rm_{1}i\rm_{2}}\langle\tilde\psi\rm_{nk}|\tilde 
p\rm_{i\rm_1}^a\rangle \langle\phi\rm_{i\rm_1}\sigma|H^{SO}(k)|
\phi\rm_{i\rm_2}\sigma'\rangle \langle\tilde p\rm_{i\rm_2}^a|
\tilde\psi\rm_{n'k}\rangle
\end{array}
\end{equation}     
where the\ spinors are chosen\ along the z-axis\ as\ 
default.\ It is also\ possible to obtain\ the eigenstates\ 
of the\ full\ spin-orbit\ Hamiltonian\ as well as the spin\
character\ along the z-axis.\ The spin character\ is\ 
defined\ as
\begin{equation}
s\rm_{mk}\equiv\langle mk|\sigma\rm_z|mk\rangle
\label{eq11}
\end{equation}
and is useful\ for analysing the degree of\ spin-orbit\ 
induced hybridization\ between spin\ up and spin down\ 
states~\cite{TO2016}. 

\section{Results\ and\ Discussion\label{sec:resdisc}}
Experimental\ refined\ x-ray diffraction structural\ 
positions were\ obtained from\ Inorganic Crystal\ 
Structure Database~(ICSD)~\cite{PP2007}.\ Structural\ 
volume optimization\ made over\ lattice range 5.0~$\rm\AA$\ 
to 6.0~$\rm\AA$,\ and obtained $a(\rm\AA)=5.5803663$,\ 
$b(\rm\AA)=7.9236074$,\ $\&$\ $c(\rm\AA)=5.6120707$,\ 
and\ $\alpha(\rm^{\circ})=90$,\ $\beta(^{\circ})=90$,\ 
$\&$\ $\gamma(\rm^{\circ})=90.015$\ which are in\ 
close\ agreement to\ experimental parameters.\ The\ 
bulk modulus\ at monoclinic\ $P{2}\rm_{1}/m$\ space\ 
group~(Fig.~\ref{fig1})\ is found to\ be\ 
$B=161.582~\rm{GPa}$,\ $\&$\ also shown\ weak\ 
ferromagnetism\ with average\ a\ remnant magnetization\ 
of 3.5$\times$10$\rm^{-6}$\ $\mu\rm_{B}$/Fe.\ 
This value is close\ to literature\ value\ 
4$\times$10$\rm^{-6}$~$\mu\rm_{B}$/Fe~\cite{ZLWCM2005}.\ 

\begin{figure}[htbp!]
\centering
\includegraphics[scale=0.9]{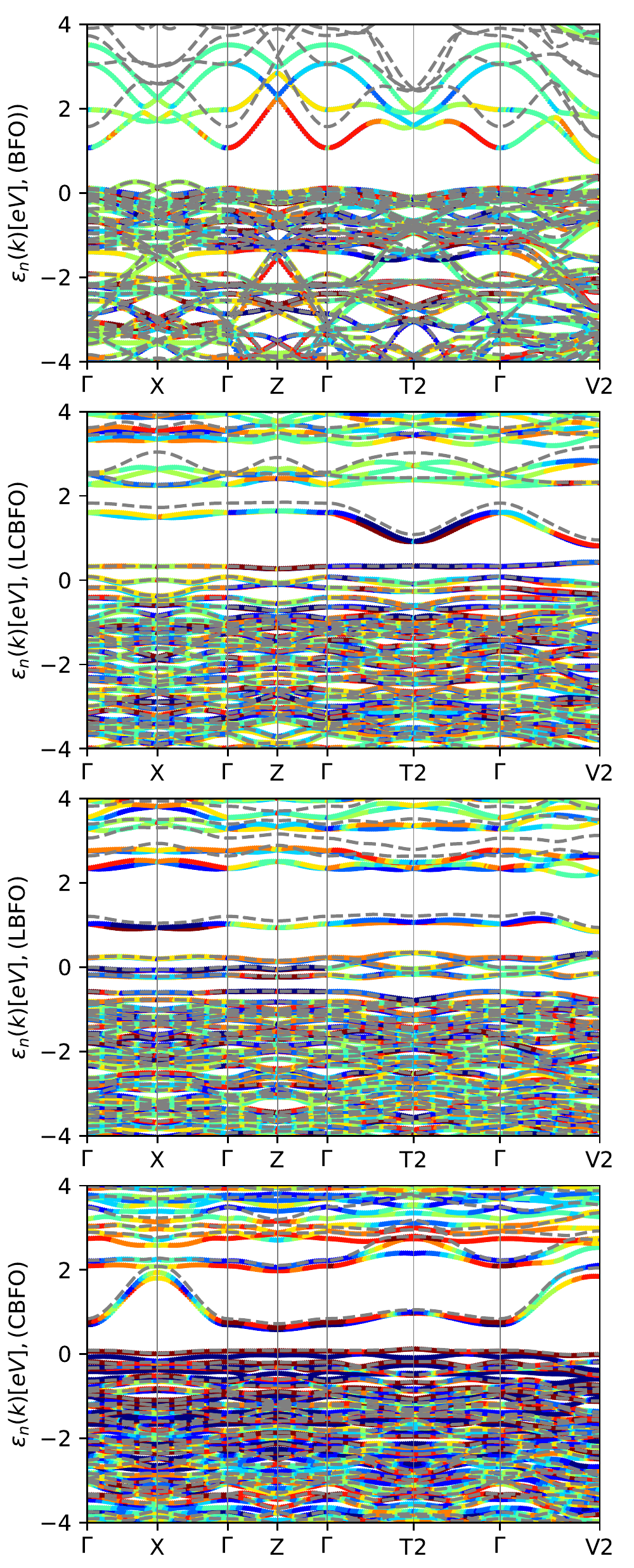}
\caption{\label{fig2}Scatter\ spin-orbit coupled\ 
band structure\ calculated using\ Eq.~(\ref{eq11}).\
The\ colors show\ the spin\ character:~blue\ 
designated as\ spin down,\ red as spin up\ 
and dashed gray\ lines as bands\ without\ 
spin-orbit\ coupling respectively.~(For\ color\ 
online,\ the reader\ is referred\ to the web\ 
version of\ this article).}
\end{figure}

Band structure\ is\ calculated in\ Fig.~\ref{fig2}\ 
using\ spin-orbit\ coupling module\ as implemented\ 
in\ the\ Gpaw\ electronic\ structure code,\ $\&$\ 
plotted over\ special k-points\ which were\ 
exported\ online\ according to\ a\ 
literature~\cite{HPKOT2017}\ by\ 
inputting\ the\ crystallographic~information\ 
file\ of $P2\rm_{1}/m$ space group.\
%%%%%%%Figure 2%%%%%%%%%%%%%%%%%%%
%%%%%%%%%%%%%%%%%%%%%%%%%%%%%%%%%%%
%%%%%%Figure 3%%%%%%%%%%%%%%%%%%%%%
\begin{figure}[h]
\centering
\includegraphics[scale=0.58]{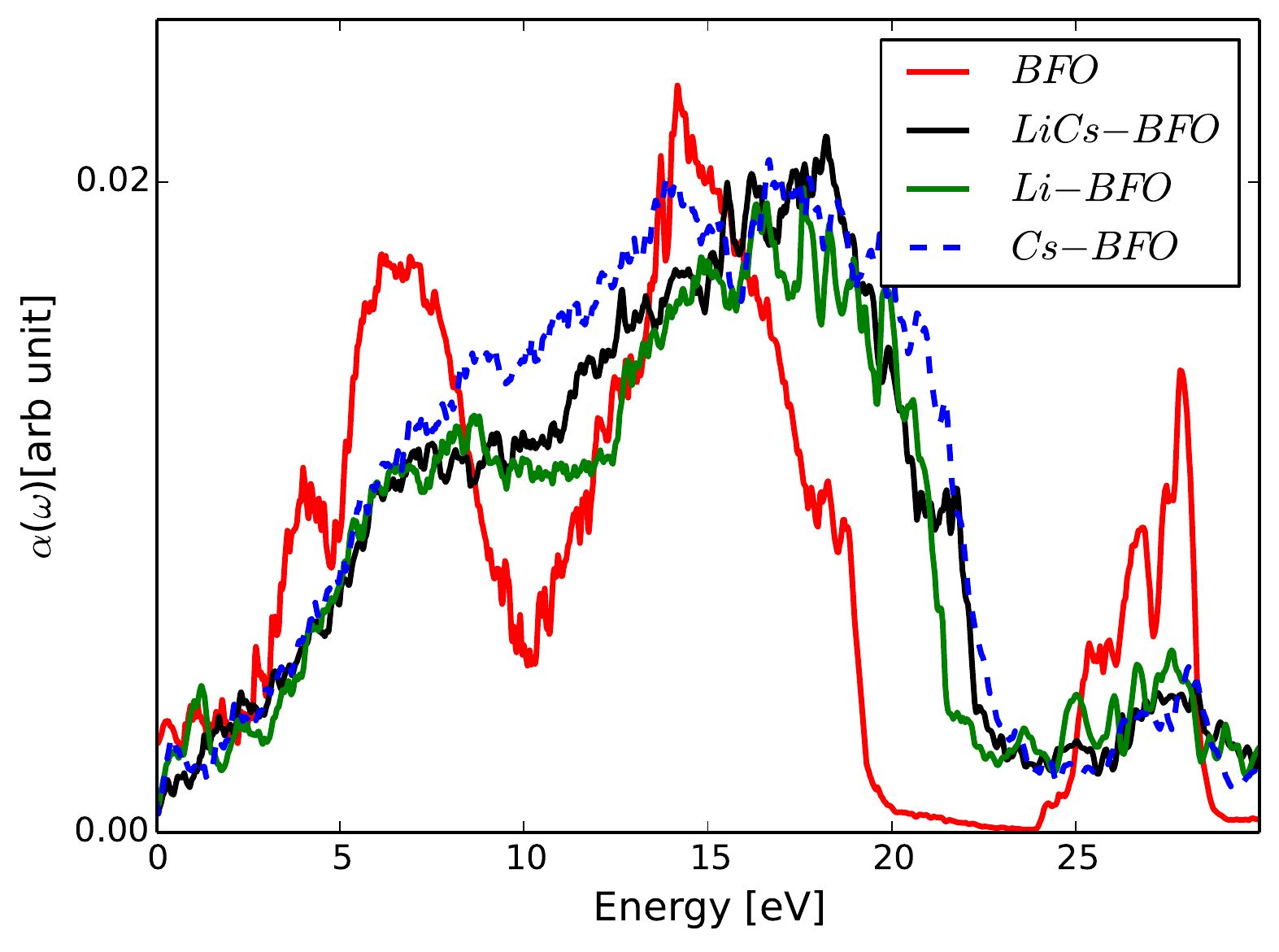}
\caption{\label{fig3}The absorption\ coefficient\ 
$\alpha(\omega)$\ for pristine\ BiFeO$\rm_{3}$\ 
in comparison to\ doped\ BiFeO$\rm_{3}$.\ 
(For\ color\ online,\ the reader\ is referred\ 
to the web\ version of\ this article).}
\end{figure}
%%%%%%%%%%%%%%%%%%%%%%%%%%%%%%%%%%% 

The calculated\ absorption\ coefficient $\alpha(\omega)$\ 
of pristine\ compared with\ doped bismuth\ ferrite\ 
are shown in Fig.~\ref{fig3}.\ The\ 
absorption\ ability\ seems\ to\ be\ reduced\ 
for\ all doped\ systems compared\ to pristine\ 
BFO in\ low\ energy-loss region.\ In addition,\ 
the optical\ bandgap\ is\ significantly\ 
decreased\ as illustrated\ in\ the\ Table.~\ref{tab1},\ 
but\ a\ dopant\ size\ also seems\ to\ have\ an\ 
effect\ on\ the\ optical\ gap~(Fig.~\ref{fig2}).\

%%%%%%%%%%%Table 1 %%%%%%%%%%%%%%%%%%%%%%%%%%%%%
\begin{table*}[!htbp]
\addtolength{\tabcolsep}{2.0mm}
\renewcommand{\arraystretch}{1.2}
\caption{Calculations\ of\ magnetization,\ 
         optical band\ gap energy,\ $\&$\ 
         magnetic\ moment\ contribution\ 
         per Fe atom,\ is\ given.\ 
         Wherever\ it\ applies\ in\ this\ 
         paper,\ BFO\ means\ pristine\ BiFeO$\rm_{3}$,\ 
         LiCs-BFO\ means\ Li~$\&$~Cs\ co-doped\ BFO,\ 
         Li-BFO\ means\ Li\ doped\ BFO,\ $\&$\ 
         Cs-BFO\ means\ Cs doped\ BFO.}

\centering
\begin{tabular}{l c c c} 
\hline\hline
{System} & {Magnetization~[$\mu\rm_{B}$]} & {Optical band gap~[eV]} &
{Moment per Fe-atom~[$\mu\rm_{B}$/Fe]}\\[0.9ex] 
\hline
BFO & {1.5$\times$10$\rm^{-6}$} & {2.13275} & {4$\times$10$\rm^{-6}$}\\
LiCs-BFO & {0.000} & {0.6855} & {0.000}\\
Li-BFO & {2$\times$10$\rm^{-6}$} & {0.7125} & {1$\times$10$\rm^{-6}$}\\
Cs-BFO & {1.532450} & {0.6746} & {0.71731}\\[1ex]
\hline
\end{tabular}
\label{tab1}
\end{table*}
%%%%%%%%%%%%%%%%%%%%%%%%%%%%%%%%%%%%%

The imaginary part,\ $\varepsilon\rm_{2}(\omega)$\ 
(Fig.~\ref{fig3})~shows\ absorption\ peaks\ 
in\ a\ photon\ energies\ range of\ 0.0-10.0~eV.\
The\ imaginary part\ is associated\ with only\ 
inter-band\ transition,\ and the\ real part\ 
$\varepsilon\rm_{1}(\omega)$\ approaches\ to\ 
zero\ at\ certain\ realms\ of\ photon\ 
energies,\ signaling a\ resonance at\ which\ 
collective\ electrons\ start\ to\ excite\ at\ 
plasma\ frequency.~
%%%%%%%%Figure 4%%%%%%%%%%%%%%%%%%%%%%%%
\begin{figure}[!htb]
\centering
\includegraphics[scale=0.28]{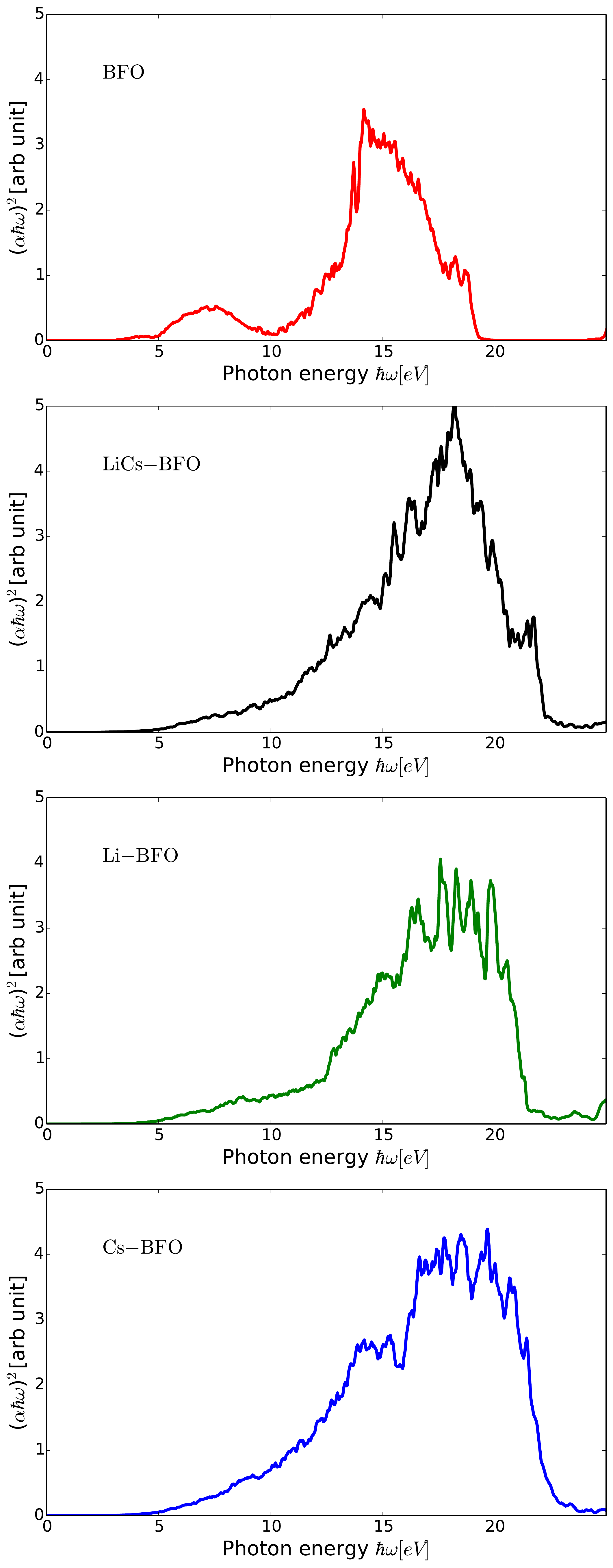}
\caption{\label{fig4}Tauc\ plots $({\alpha}{\hbar}{\omega})\rm^{2}$\ 
        vs.\ photon energy\ ${\hbar}{\omega}$\ for direct\ 
        band gap E$\rm_{g}^{o}$\ extrapolation\ or estimation\ 
        of\ pristine\ BFO,\ Li\ $\&$\ Cs co-doped BFO,\ 
        Li doped BFO,\ $\&$,\ Cs doped\ BFO in\ respective\ 
        order of\ from top\ to bottom.}
\end{figure}
%%%%%%%%%%%%%%%%%%%%%%%%%%%%%%%%%%%%%
%%%%%%%%Figure 5%%%%%%%%%%%%%%%%%%%%%%%%
\begin{figure}[!htb]
\centering
\includegraphics[scale=0.485]{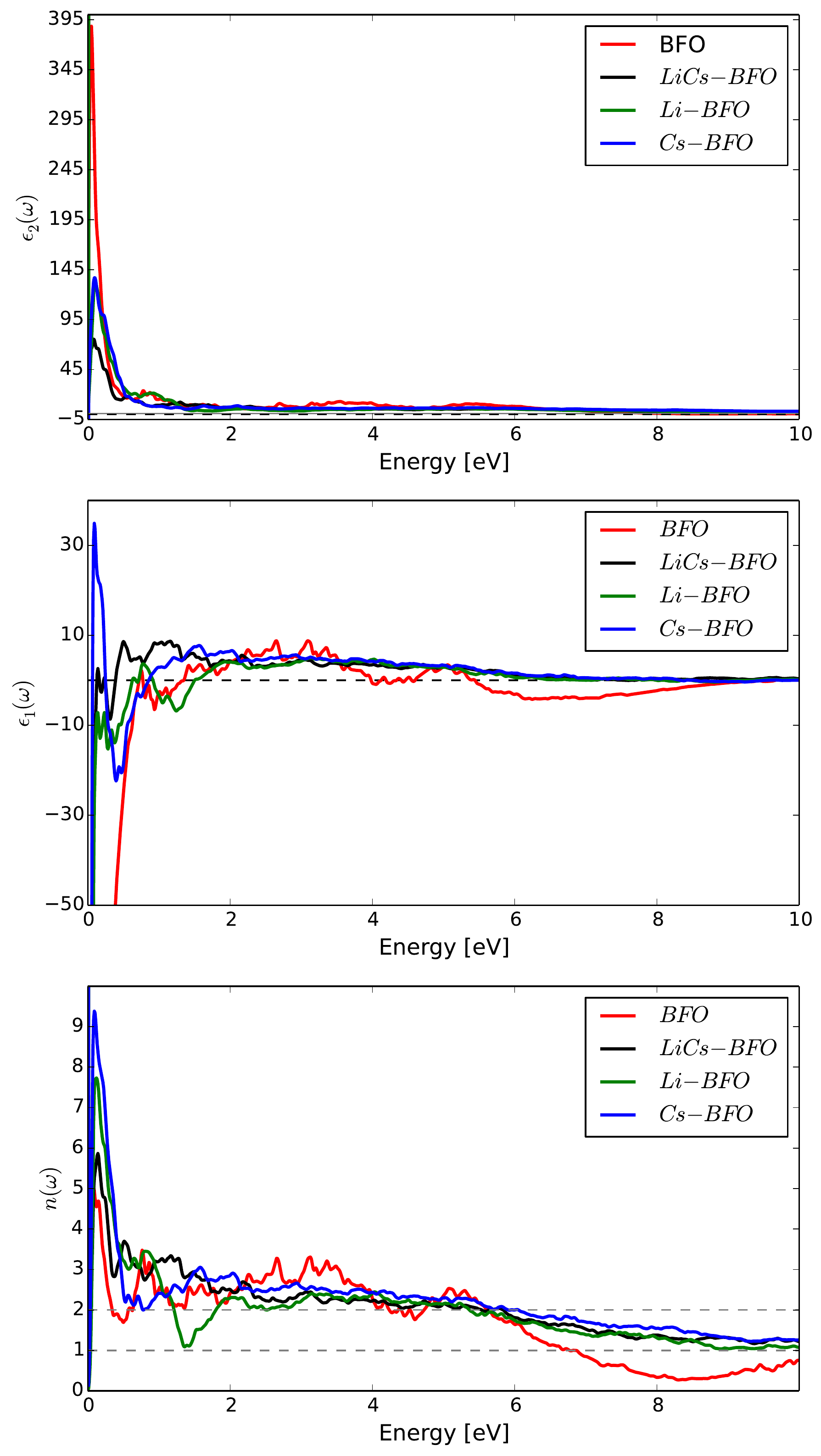}
\caption{\label{fig5}The calculated real part\ 
$\varepsilon\rm_{1}(\omega)$\ and imaginary part\ 
$\varepsilon\rm_{2}(\omega)$ of\ dielectric function\ 
$\varepsilon(\omega)$ $\&$\ refractive index\ 
$n(\omega)$\ in pristine\ BiFeO$\rm_{3}$,\ compared\ 
with doped\ BiFeO$\rm_{3}$.\ (For\ color\ 
online,\ the reader\ is referred\ to the web\ 
version of\ this article).}
\end{figure}
%%%%%%%%%%%%%%%%%%%%%%%%%%%%%%%%%%%%%
It\ looks\ from\ Fig.~\ref{fig2}\ that\ 
a\ congested bound\ of bands near\ the\ 
Fermi\ level of\ valence band\ seems\ to\ 
oscillate with\ plasma frequency.\ 
From\ Fig.~\ref{fig4},\ the\ optical\ 
bandgaps\ presented\ in\ the\ Table~\ref{tab1}\
are\ calculated\ using\ Tauc~relation~(Eq.~\eqref{eq8}),\
\&\ according\ to\ a\ tangent\ line\ 
curve\ fitting\ approach\ adopted\ from\ 
a\ literature~\cite{PCHOT2017}.\ 

We also\ notice\ from\ Figs.~\ref{fig4}~$\&$~\ref{fig5}\ 
that\ absorption\ peaks\ occur\ over\
a\ wider\ ranges\ of\ photon\ energies\ 
by\ doping\ the\ BFO,\ $\&$\ that\ a\ 
decrease\ in\ the\ imaginary\ part\ 
$\varepsilon\rm_{2}(\omega)$\ at\ higher\ 
photon\ energy\ potentially\ indicates\ 
a\ decline\ in\ absorption of\ photon\ 
energy\ to\ bound\ electron of\ valance\ 
band.\ Thus,\ the\ excitation\ of the\ 
plasmons\ by light\ is\ demonstrated\ 
by a\ decrease of\ the reflectivity\ 
$R(\omega)$\ of\ the light\ at plasma\ 
frequency\ because\ of\ photon\ energy\ 
is\ consumed\ up by\ oscillating\ bound\ 
electrons.\ The refractive\ index as\ 
indicated in\ Fig.~\ref{fig5}\ show\ 
that\ $n(\omega)~{\rightarrow}~1$,\ 
at\ high\ photon\ energies~(${\sim}$~10-15~eV).\ 
This\ could\ potentially\ imply\ that light\ 
can\ pass\ through\ the\ material without\ a\ 
phase\ change,\ which\ is a\ novel\ 
characteristics\ of\ photonic\ crystals.\ 
Such\ a\ claim\ also\ agrees\ with\ experimental\ 
$\&$\ theoretical\ concepts\ explained\ in\ 
literature~\cite{JS2011, LE2017}. 
%%%%%%%%Figure 6%%%%%%%%%%%%%%%%%%%%%%%%
\begin{figure}[!htb]
\centering
\includegraphics[scale=0.365]{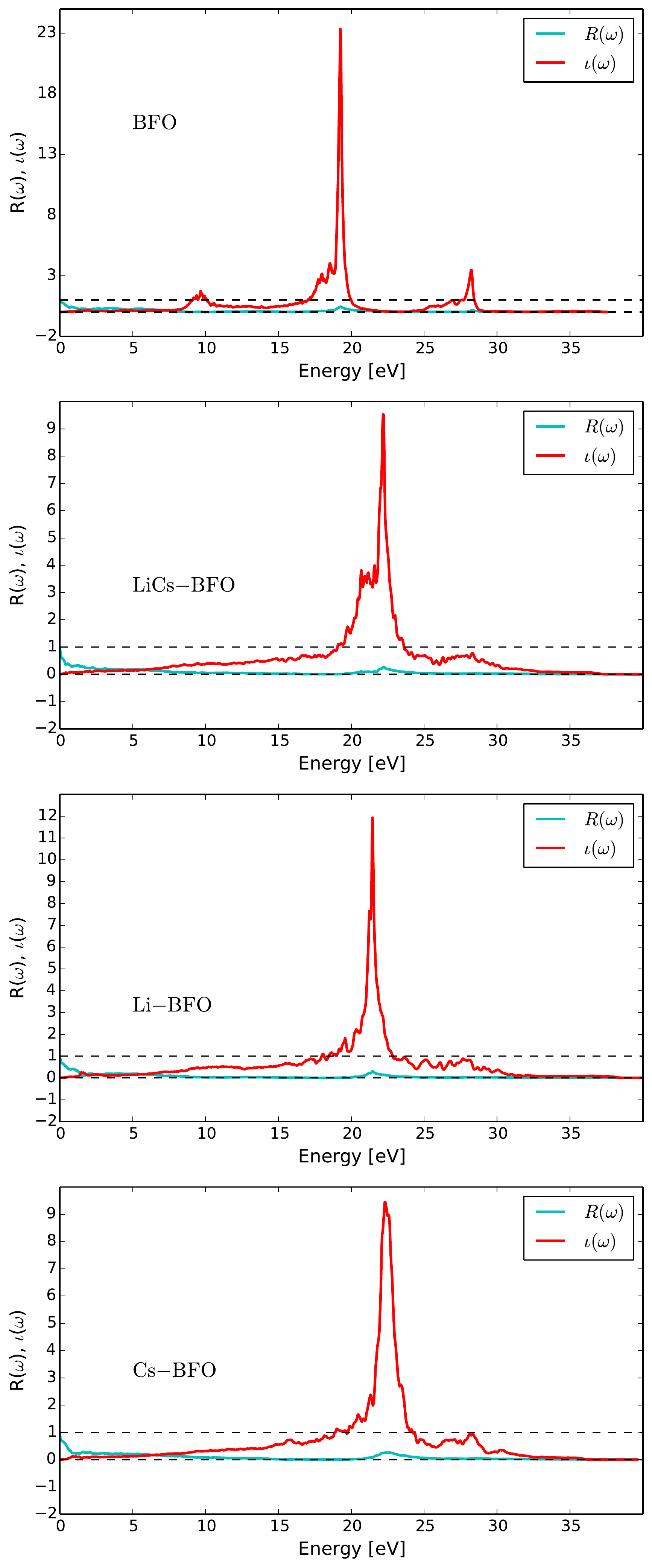}
\caption{\label{fig6}The energy loss,\ 
$\iota(\omega)$ $\&$ reflectivity,\ $R(\omega)$\  
for pristine\ BiFeO$\rm_{3}$\ in comparison\ to\ 
doped BiFeO$\rm_{3}$.\ (For\ color\ 
online,\ the reader\ is referred\ to the web\ 
version of\ this article).}
\end{figure}
%%%%%%%%%%%%%%%%%%%%%%%%%%%%%%%%%%%%%

As shown\ in Fig.~\ref{fig6},\ plasmons move\ 
in a\ periodic potential\ and unlike ideal plasmons,\ 
and pristine BFO,\ had three peaks at 10 eV,\ 19~eV\ 
$\&$ 27~eV.\ Li\ $\&$\ Cs\ co-doped BFO had three\ 
peaks at 21~eV,\ 24~eV,\ $\&$ 27~eV.\ Li doped BFO\ 
had two peaks\ at 21~eV,\ $\&$ 28~eV.\ Cs doped\ 
BFO had three\ peaks at 15~eV,\ 26~eV,\ $\&$ 28~eV.\
The energy-loss\ may show plasmon\ excitation\ 
in region of\ about\ 10-30~eV often\ during\ 
inter-band\ transition of\ single electron.\ 
However,\ high energy-loss\ region\ is\ highly\ 
associated\ with excitation\ of bound electron\ 
with multiple\ peaks.\ These bound\ electron\ 
significantly\ increase\ photocurrent\ generation\ 
which\ enormously enhance\ photovoltaic effect.\ 
It is been\ noted that\ two\ plasmons\ of pristine\ 
BFO\ interact\ via electron-hole\ pair resulting\ 
in biplasmons\ generation due\ to Cs\ doping.\ 
However,\ Li doping\ affected the\ amount\ of\ 
energy\ of the\ two peak\ points plus\ a\ very\ 
small\ peak at 30 eV.\ Therefore,\ Cs size\ 
had effect on\ biplasmons generation\ in\ 
comparison to\ the Li size.\ In some cases,\ 
plasmons\ width\ $\&$\ peak explain\ the phase\ 
changes.\ We can\ infer that\ size of the\ 
dopant atom has\ remarkable impact\ on both\ 
number\ $\&$\ width of\ plasmons resulting\ 
in\ enhanced photovoltaic\ property.\ Hence,\ 
co-doping\ of\ Li\ $\&$\ Cs in\ a\ Barium\ 
doped\ bismuth ferrite\ seems\ to\ have\ a\ 
potential\ for\ improved\ efficiency\ as\ a\ 
photoelectron\ capturing device.\

\section{Conclusion\label{sec:conc}}
The\ investigations\ can\ be\ summarized\
as\ follows.\
\begin{itemize}
\item With\ doping,\ absorption\ coefficient\
has\ widened\ resulting\ to\ a\ decreased\ 
optical\ bandgap\ $\&$\ hence\ shows\ a\ 
potential\ for\ an\ increased\ solar\ cell\ 
property.\ 
\item A\ squared\ absorption\ coefficient\ 
peak\ shows\ an increase\ in width of\ 
by about 4~eV\ towards\ a\ realm of\ 
higher\ photon\ energies\ in\ the\ case\ 
of\ doped\ BFO.\ This\ is\ believed\ to\ 
be\ a\ sign\ of\ an\ increased\ solar\ 
cell\ activity.\
\item Dielectric\ functions\ seem\ to\ 
have\ large\ values\ at\ zero\ photon\ 
energies,\ but\ these\ values\ seem\ to\ 
be\ reduced\ significantly\ by\ an\ 
introduction\ of\ a series\ of\ photon\ 
energies\ of\ up~to\ ${\hbar}{\omega}$=10~eV.\
For\ photon\ energies\ of 10-15 eV,\ 
$n(\omega)~{\rightarrow}~1$.\   
\item The\ crossing\ of\ the\ imaginary\ 
part\ of\ the\ dielectric\ function\ 
$\varepsilon\rm_{1}(\omega)$\ of\
a\ zero\ values\ at\ multiple\ realms\ 
of\ the\ photon\ energies\ $\hbar\omega$\ 
indicates\ a\ potential\ importance\ of\ 
the\ system\ in\ applications\ based\ 
on\ photonic\ conductivity.\
\item The\ systems'\ $\varepsilon\rm_{1}(\omega)$,\ 
$\varepsilon\rm_{2}(\omega)$,\ $\&$\ $n(\omega)$\ 
seem\ to\ be\ reduced\ rapidly\ with\ 
introduction\ of\ photon\ energies,\
while\ a\ major\ contributor to\ the\ 
magnitude\ of\ $n(\omega)$\ seems\ to\ 
be\ $\varepsilon\rm_{2}(\omega)$.\  
\item Expanded\ width\ of\ energy\ loss\ 
peaks\ happen\ with\ doping\ $\&$\ an\ 
occurrence\ of\ such\ peaks\ at\ higher\ 
photon\ energies\ indicates\ a\ potential\ 
importance\ for\ applications\ in\ 
a\ plasmonic\ effect.\     
\end{itemize}
\section*{Disclosure\ statement}
The authors declare\ that there\ is no conflict\ 
of interest.

\section*{Acknowledgements}
We are grateful\ to the Ministry of\ Science\ 
and\ Higher\ Education\ of Ethiopia for\ financial\ 
support.\ The\ authors also\ acknowledge\ the\ 
Department of\ Physics at\ Addis\ Ababa\ University\ 
and the\ International\ Science Program,\ Uppsala\ 
University,\ Sweden,\ for\ providing computer\ 
facilities.\ 

\newpage
\section*{ORCID\ iDs}
K.N.\ Nigussa.\\
\url{https://orcid.org/0000-0002-0065-4325}.
\bibliographystyle{elsarticle-num}
\bibliography{refs}
\end{document}